# A Heuristic Scheduling Scheme in Multiuser OFDMA Networks

Zheng SUN, Zhiqiang HE, Ruochen WANG, Kai NIU

School of Information Engineering, Beijing University of Posts and Telecommunications, Beijing, China

zhengs.bupt@gmail.com

*Abstract*—Conventional heterogeneous-traffic scheduling schemes utilize zero-delay constraint for real-time services, which aims to minimize the average packet delay among real-time users. However, in light or moderate load networks this strategy is unnecessary and leads to low data throughput for non-real-time users. In this paper, we propose a heuristic scheduling scheme to solve this problem. The scheme measures and assigns scheduling priorities to both real-time and non-real-time users, and schedules the radio resources for the two user classes simultaneously. Simulation results show that the proposed scheme efficiently handles the heterogeneous-traffic scheduling with diverse QoS requirements and alleviates the unfairness between real-time and non-real-time services under various traffic loads.

## I. Introduction

Recent years have seen a rapid growth of high speed wireless communication technologies. As the downlink access data rate increases, future wireless networks are expected to provide scheduling for heterogeneous services with different quality of service (QoS) demands [1]. Many papers addressing this issue have been published, such as the work in [2], which introduces a method to inherit utility function that maps different service rates to fairness among users. In [3], the authors define users' satisfaction factor to quantify QoS for different service types and present various formulations therein. Since ideas in the two works are based on defining specific mapping functions from physical parameters to service qualities, the accuracy and sensitivity may not be guaranteed. Additionally, the work in [4] shows a modified exponential scheduling scheme, which accounts multiple QoS factors in radio resource assignment, such as maximum acceptable packet drop probability, status of packet queues, the degree of QoS satisfaction and maximum tolerable delay. This scheme behaves in a proportional fairness manner and deals with both non-real-time and real-time sessions. In [5], the objective of maximizing the minimum throughput rate is considered, and [6] extends this framework to support dynamic user population.

However, in traditional schemes for heterogeneous services as mentioned above, we see that the scheduling schemes for real-time and non-real-time users are always conducted sequentially and are under *zero-delay* constraint, i.e. the system first schedules radio resources (i.e. time slots, subcarriers and codes in TDMA, OFDMA and CDMA systems, respectively) to minimize the average packet delay for real-time services, and then allocates the leftover resources to best-effort users. This strategy works well in heavy-load networks to guarantee QoS provision for real-time users, but it also leads to a dilemma that when it is adopted in light or moderate load networks, the real-time services usually get over-satisfied while the leftover resources for non-real-time ones is very little. This problem has also arisen and been analyzed in wired networks such as ATM, but many research shows that the scheduling algorithms suitable in wired networks may not still work efficiently, or even feasibly, in wireless scenarios, where the users have to experience time-varying channel conditions and where the data access rates become extremely fluctuating.

To solve this problem, in this paper we try to design a heuristic scheduling scheme for wireless OFDMA system that combines the two service types together. Motivated from the idea that in traditional scheduling schemes radio resources for real-time services are assigned to different users based on (weighted) QoS priority comparison [2]-[4], the proposed scheme innovatively assigns non-real-time users an adaptive scheduling weight as well, which will be compared with the scheduling priorities of real-time users when assigning subcarriers. The inspiration behind is straightforward. If the scheduling weight for non-real-time users is set to zero, the proposed scheme is the same as traditional ones. And if the system adaptively increases the scheduling weight, more and more radio resources would be *borrowed* from real-time users to non-real-time users based on scheduling priority comparison. In the numerical analysis it will be shown that if the value of the scheduling weight for non-real-time users is set properly, the system guarantees to only provides the real-time users with their *necessary* packet delay constraint (not zero-delay), and the scheduler reduces the unfairness between the two service classes by allocating as much radio resources (i.e. subcarriers) to non-real-time services as possible while guaranteeing the necessary QoS provision for real-time services.

The rest of this paper is organized as follow. In Section II, we describe the channel and traffic model, and formulate the optimization problem. In Section III, our idea and the proposed scheduling scheme are presented. Simulation conditions and results are shown in Section IV. And Section V provides a summary and concludes the work.

## II. System Description

In this section, we briefly describe the studied network and traffic models, and formulate the cross-layer scheduling optimization problem. Some important symbols utilized to analyze the system performance are defined in Table I. Other useful symbols will be defined when they first appear in context.

### A. Network Scenario Description

In this paper, we consider a single-cell downlink OFDMA system with one Base Station (BS) and $N$ Mobile Stations (MS's, also referred as users). Inter-cell interference is not taken into consideration. The total system bandwidth $W$ is divided into $K$ subcarriers, and each subcarrier has a bandwidth of $\Delta f = W / K$.

This research is sponsored by Project 60772108 supported by National Natural Science Foundation of China and National Basic Research Program of China (973 Program), 2007CB310604

TABLE I
SYMBOL DEFINITIONS

| Symbol | Definitions |
| --- | --- |
| $D$ | set of subcarriers. |
| $D_{QoS(BE)}$ | set of subcarriers assigned to QoS (BE) traffic (or equivalently, QoS (BE) users), in a subcarrier allocation. |
| $\{D_{QoS}, D_{BE}\}$ | one specific subcarrier allocation |
| $U$ | set of all users |
| $U_{QoS(BE)}$ | set of QoS (BE) users |
| $U_i$ | user $i$ in a specific user set |
| $D_i$ | set of subcarriers assigned to user $i$ in a specific user set |

The OFDMA scheduling is time-slotted, and the length of each time slot is $T_s$. Although the users experience time varying wireless channel, we assume the channel condition of each user stays constant within a single time slot. The channel state information (CSI) is assumed to be perfectly estimated, and reliably fed back from users to the base station, so that the scheduler can perform subcarrier scheduling based on the CSI once per time slot. Let $H_{ik}$ denote the channel frequency response at subcarrier $k$ of user $i$. The SNR of user $i$ at this subcarrier can be expressed as $\eta_{ik} = \Gamma \cdot |H_{ik}|^2 P_{ik} / N_{ik}$, where $\Gamma = -\ln(5\text{BER})/1.5$, $P_{ik}$ is the transmit power for user $i$ at subcarrier $k$, and $N_{ik}$ is the noise power density. In this paper, we assume that the transmit power is uniformly distributed over the entire available frequency band and set to unit value. Let $r_{ik}$ denote the channel capacity of user $i$ at subcarrier $k$, then $r_{ik} = \log_2(1 + \eta_{ik})\Delta f$ (bit / s). Defined as the sum of channel capacities of all its scheduled subcarriers, the total achievable throughput of one user in a time slot is given by $\sum_{k \in D_i} r_{ik}$.

### B. User Type Classification

The various services are classified into two classes based on their delay tolerance. The QoS service represents the ones which are delay sensitive, and requires the scheduler to provide a certain data rate on downlink. This type of application includes many high-speed real-time downlink data services that are widely studied nowadays, such as *Video on Demand* (VOD), packet-switched voice, FTP service. In contrast, the other class corresponds to the best-effort (BE) services conducting more elastic applications such as file transfer and e-mail. This kind of services can adjust their data rates gradually and are often delay tolerant. The users in the system are divided into QoS and BE users according to the traffic classes their services belong to.

### C. Optimization Problem Formulation

The optimization problem for subcarrier scheduling can be expressed as follows:

$$\underset{D_i, i \in U}{\text{maximize}} \sum_{i \in U} A_i \sum_{k \in D_i} r_{ik} \quad (1)$$

subject to $\bigcup_{i \in U} D_i = D; D_j \bigcap D_i = \varnothing, i \neq j, \forall i, j \in U$,

where $A_i$ denotes the scheduling priority representing the QoS satisfaction of user $i$. Notice that in traditional scheduling schemes, only real-time users have such scheduling priorities, since the schemes are designed only with the aim to guarantee required QoS provision for real-time users, and take not much consideration about non-real-time users such that the rate allocation for these users is conducted in a best-effort pattern. Motivated from the idea that in an ideal subcarrier allocation, the scheduler should not only provide the delay-control for QoS users, but also optimize the overall throughput for BE users, we formulate a dual-objective programming problem for OFDMA system as

$$\begin{cases} \underset{D_i, i \in U_{QoS}}{\text{maximize}} \sum_{i \in U_{QoS}} A_i \sum_{k \in D_i} r_{ik} \\ \underset{D_i, i \in U_{BE}}{\text{maximize}} \sum_{i \in U_{BE}} \sum_{k \in D_i} r_{ik} \end{cases} \quad (2)$$

To make (2) more tractable, we introduce a weighted parameter $\lambda \in R^+$ to balance the overall allocation priorities between the two user types, so that (2) is transformed into an equivalent programming problem as follows.

$$y = \underset{D_i, i \in U}{\text{maximize}} \left\{ \sum_{i \in U_{QoS}} A_i \sum_{k \in D_i} r_{ik} + \lambda \cdot \sum_{i \in U_{BE}} \sum_{k \in D_i} r_{ik} \right\} \quad (3)$$

subject to $\bigcup_{i \in U} D_i = D; D_j \bigcap D_i = \varnothing, i \neq j, \forall i, j \in U$.

In the scheduling comparison based on (3), the scheduling priority $A_i$ for QoS user $i$ already takes QoS provision into consideration. Since $\lambda$ in (3) is equivalent to $A_i$, we can consider $\lambda$ as the scheduling priority for the BE users. Therefore, in light or moderate load networks a method of increasing the access data rate for BE users and alleviating the unfairness between the two service types is to adaptively (and gradually) increase the value of $\lambda$ in every time slot, so that the BE users will occupy resources (i.e. subcarriers) of the QoS users based on priority comparison. In practical designing, there are two remaining problems to solve: firstly, if $\lambda$ is given for a time slot, how to assign subcarriers to maximize (3); secondly, how to adaptively update $\lambda$ during different time slots so as to provide as many subcarriers to BE users as possible while keeping the average packet delay for QoS users lower than a certain threshold. In what follows, we will focus on tackling these problems.

### III. DYNAMIC SUBCARRIER ALLOCATION

As mentioned in Section II.C, if $\lambda$ is predetermined at the beginning of a time slot, the problem in (3) is similar to the traditional scheduling problem only with difference that the BE users are also assigned scheduling priorities (i.e. $\lambda$) and join the scheduling comparison with the QoS users. In this section, we will firstly present an optimal scheduling algorithm with given $\lambda$, and then propose a heuristic for updating $\lambda$.

### A. Optimal Scheduling Algorithm with Given $\lambda$

The algorithm comprises two levels. The higher traffic-level scheduling is implemented between traffic classes. We conclude with the following theorem consisting of dual parts, which account for the QoS and the BE traffic, respectively.

*Theorem 1:* **In traffic-level scheduling, given $\lambda$, the formula (3) achieves maximum if and only if:**

**(Qos traffic) If allocating subcarrier $p$ to user $k$, $k \in U_{QoS}$, it satisfies** $\forall l \in U_{BE}$, $r_{kp}/r_{lp} \geq \lambda/A_k$.

**(BE traffic) If allocating subcarrier $q$ to user $m$, $m \in U_{BE}$, it satisfies** $\forall j \in U_{QoS}$, $r_{jq}/r_{mq} \leq \lambda/A_j$.

*Proof*: Since the proofs of the two parts are similar, we only show the proof of the QoS traffic. Let $\{D_{QoS}^*, D_{BE}^*\}$ be the optimal subcarrier allocation which maximizes (3). Randomly choose one subcarrier $p$ from $D_k$, $k \in U_{QoS}$, move it from $D_{QoS}^*$ to $D_{BE}^*$ and reallocate it to $l$, $l \in U_{BE}$. Since (3) is maximized, the change would not increase it, which means

$$\sum_{i \in U_{QoS}} A_i C_i - A_k r_{kp} + \lambda \left( \sum_{i \in U_{BE}} C_i + r_{lp} \right) \leq \sum_{i \in U_{QoS}} A_i C_i + \lambda \sum_{i \in U_{BE}} C_i,$$

which is equivalent to $r_{kp}/r_{lp} \geq \lambda/A_k$. (4)

It implies that if a user $l$, $l \in U_{BE}$ satisfies $r_{kp}/r_{lp} < \lambda/A_k$, then moving $p$ from $D_{QoS}$ to $D_{BE}$ can definitely optimize (3). Therefore the optimal subcarrier allocation must satisfy $\forall l \in U_{BE}$, $r_{kp}/r_{lp} \geq \lambda/A_k$.

The sufficiency of Theorem 1 is straightforward. From the subcarriers' point of view, the optimization problem in (3) can be viewed as a sum of components, each of which represents a subcarrier's assignment being independent to each other. This is to say, if every subcarrier's assignment maximizes its own component, the overall performance achieves optimal. Therefore Theorem 1 is sufficient. ∎

When traffic-level scheduling finishes, the lower user-level scheduling determines how to assign subcarriers to every individual user of the traffic class decided in the traffic-level. Since by Theorem 1 the subcarriers have been classified into two groups corresponding to service types, the scheduling for the two service types could be considered as two separate scheduling problems such that the $A_i$'s and the $\lambda$'s behave as the scheduling priorities for the QoS users and the BE users, respectively.

*Theorem 2:* Let us denote $r'_{ip} = A_i r_{ip}$ (or $r'_{ip} = \lambda r_{ip}$) as the weighted data rate for QoS (BE) user $i$ at subcarrier $p$. In user-level scheduling, given $\lambda$, if the subcarrier allocation makes (3) optimal, the subcarriers must be assigned to the QoS (BE) users with the greatest weighted data rates.

*Proof*: As traffic-level scheduling finishes, the scheduling for QoS users can be rewritten as

$$f = \max_{u(k), k \in D_{QoS}} \left\{ \sum_{k \in D_{QoS}} r'_{u(k)k} \right\}, \quad (5)$$

where $u(k), k \in D_{QoS}$ is the mapping from subcarrier $k$ to user $u(k)$. Notice that (5) is a strictly increasing function to $r'_{u(k)k}$ ($k \in D_{QoS}$), and every subcarrier's assignment is independent to each other. Hence when (5) achieves maximum, every subcarrier should be assigned to the QoS user with the greatest weighted data rate to it.

Similarly, we can prove the optimal scheduling for BE users by substituting $A_i$ by $\lambda$. ∎

### B. Dynamic Heuristic for Updating $\lambda$

During the discussion above, it has assumed the value of $\lambda$ is already determined in every time slot. In this section, we will develop a heuristic for updating $\lambda$ during different time slots. Considering $\lambda$ as the scheduling priority for BE users, we notice that $\lambda$ differs from $A_i$ by means that $A_i$ is changeable for every user since it is determined by QoS requirements, while $\lambda$ is identical for all BE users and is evaluated arbitrarily in every time slot. This means that in optimization of (3), the relative priority relations between QoS users, i.e. $A_i$'s, do not change after involving $\lambda$. Therefore, when the network is in light or moderate system load, the scheduler can gradually increase $\lambda$ to make the allocation threshold $r_{kp}/r_{lp} \geq \lambda/A_k$ ($k \in U_{QoS}$, $l \in U_{BE}$) in Theorem 1 get harder to hold, so that the BE users will get more subcarriers based on priority comparison. As the average packet delay of QoS users gets close to some certain threshold, the increase stops. And when the network is overload, in order to firstly guarantee the QoS requirement for QoS users, the scheduler may decrease $\lambda$ to consequently raise the relative scheduling priority of QoS users, so that more subcarriers will be assigned to QoS users.

In summary, a heuristic based on the idea mentioned above is described in Table II. In every time slot, the system at first executes the heuristic to determine the value of $\lambda$ based on the

TABLE II
HEURISTIC ALGORITHM DESCRIPTION

| | |
|---|---|
| Initialization: | |
| Step 1. | Set $\Delta\lambda = \Delta\lambda_0$, $N_1$, $N_2$, and $\varepsilon$. |
| Step 2. | Measure $\overline{A}$ as the average of $A_i$'s of all real-time users. Set $\lambda = \overline{A}/N_1$. GOTO Updating Process. |
| Updating Process: | |
| Step 1. | Measure the average packet delay $\overline{d}$ of real-time users; |
| Step 2. | Compare $\overline{d}$ and $\overline{d}_{\max}$, i.e. the maximum packet delay of real-time users; <br> IF $\overline{d} \leq \overline{d}_{\max}$ : $\lambda = \lambda + \Delta\lambda$ ; <br> ELSEIF $\overline{d} \leq \overline{d}_{\max} \cdot (1+\varepsilon)$ : <br> $\quad \lambda = \lambda - \Delta\lambda \times 2$, and $\Delta\lambda = \Delta\lambda/N_2$. <br> Go back to Step 1. <br> ELSE : GOTO Initialization. <br> ENDIF |

Remarks:
1). $N_1$, $N_2 \in Z^+$ are predefined constant. $\varepsilon \in (0,1]$ is the relaxation factor of the maximum packet delay of real-time users.
2). The Initialization runs only once in the first time slot unless called in the Updating Process.
3). Updating Process runs at the beginning of every time slot.

QoS information, i.e. $A_i$, and then allocate subcarriers using the algorithm mentioned in Section III.A.

## IV. SIMULATION RESULTS

In this section, we design a single-cell scenario simulation with the following simulation conditions.

### A. Simulation Conditions

In simulation, the system bandwidth is assumed to be 1.024 MHz, and there are totally 256 subcarriers. The cell radius is 1 km and the path loss is calculated by $38.4 + 20\log_{10} d$[dB], where $d$ (m) is the distance between a user and the base station. Shadowing is assumed to be lognormally distributed with mean 0 dB and standard deviation 8 dB. Users are uniformly distributed in the cell, and every user moves at an average speed of 20 m/s. The standard deviation of user speed is 2.24 m/s. Each user is dedicated to one session of a specific traffic type. The scheduling performs every 0.125ms. The transmit power from the base station is fixed to 43dBm, and the thermal noise power is -108dBm. The achievable coding rates are {1/2, 2/3, 3/4, 7/8}. The selective modulation schemes are QPSK, 16QAM, 32QAM, and 64QAM. Three traffic types are considered: VoIP, video streaming (STR) and BE traffic. The VoIP traffic is adopted according to the ON/OFF model in [10]. The average durations of ON and OFF periods are 1.0s and 1.5s respectively. We assume within each ON interval, the voice data rate is 32Kbps, and the lifetime of a packet is 80ms. The streaming traffic is according to the model in [11]. The duration of each state is exponentially distributed with mean 160ms. The data rate of each state is in a truncated exponential distribution where the data rate range is from 64 to 256 with an average value of 180Kbps, and the maximum packet delay is 1s. In order to simulate the maximum performance of the BE traffic, we apply a full-buffer model, so that the maximum throughput for the best effort services can be obtained. During the simulation, we inherit the exponential (EXP) scheduling scheme proposed in [7], which defines the scheduling priority for QoS users $i$ as

$$A_i = \mu_i \exp\left(\left(\mu_i H_i - \overline{\mu H}\right) \Big/ \left(1 + \sqrt{\overline{\mu H}}\right)\right), \quad (6)$$

where $\mu_i$ equals $-\log\left(P_{D,i}^{\max}\right)/T_{\max,i}$. $P_{D,i}^{\max}$ is the maximum acceptable packet drop probability of user $i$, $T_{\max,i}$ is the initial lifetime of each packet, $H_i$ is the head-on-line delay, and $\overline{\mu H}$ is the average of $\mu_i H_i$ of all users. For each experiment, 100 simulation runs are averaged to estimate the scheduling result, and each run is executed over $2 \times 10^6$ time slots.

### B. Simulation Results

#### 1) Increase of STR users

In this simulation, we increase the number of STR users from 4 to 20, and fix the number of VoIP and BE users to be 10 and 20, respectively. The maximum packet delay $\overline{d}_{\max}$ of VoIP service is 0.5s. Since the video streaming sessions have high access data rates, i.e. 180 kbps for average, both the moderate load and congested situations are depicted in Figure 1 and Figure 2. The 14-user case is a turning point. When the user is fewer than 14, the network is less-congested, and both the proposed and the EXP schemes show good delay-control for QoS users. The delay of VoIP users in the proposed scheme is a little lower than predefined maximum delay threshold (i.e. 0.5s) and the throughput of BE users is about 55% more than that of the EXP scheme. (Note that the gain in throughput of BE users is highly relevant to the predefined maximum delay threshold $\bar{d}_{\max}$ and the user numbers.) After the network gets congested, from 14 to 20, the delay of the STR users in both the schemes has a dramatic rise due to lack of subcarrier resources per user. Notice that while the number of the STR user increases, the throughput of the proposed scheme drops down in a faster speed than that of the EXP scheduling. This is because when the packet delay of VoIP users is about to exceed the threshold, the proposed scheme would decrease $\lambda$ to guarantee the delay-control for the QoS users before scheduling subcarriers to the BE users. Therefore the throughput of the BE users is sacrificed to compensate for the loss of transfer delay for QoS users. This is consistent with the predefined function of $\lambda$ and the heuristic algorithm.

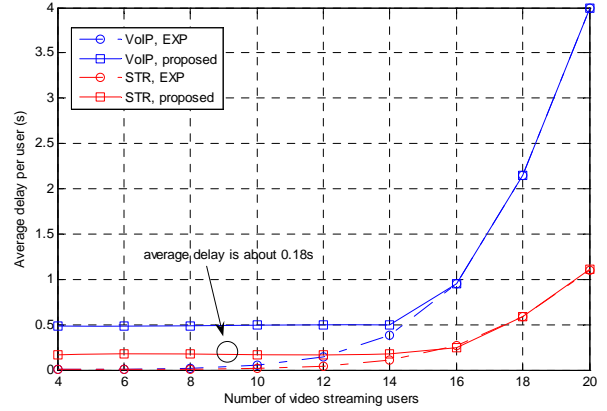

Fig. 1. Average packet delay of VoIP traffic and STR traffic when STR user number increases.

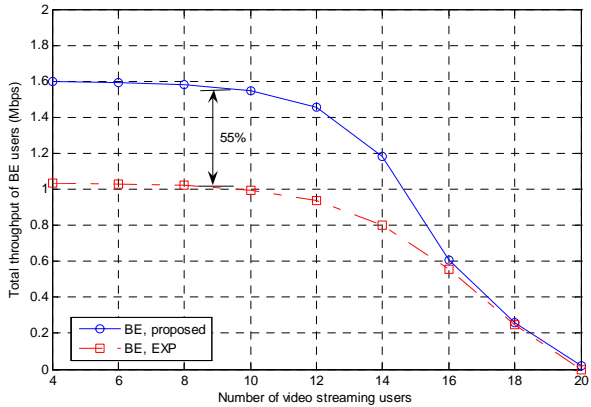

Fig. 2. Total throughput of BE users when STR user number increases.

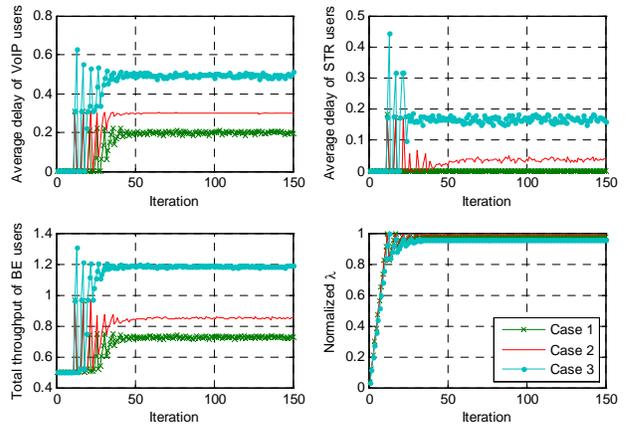

Fig. 3. Convergence of the heuristic for updating $\lambda$.

*2) Dynamic updating of $\lambda$*

In this part, we design experiments to check the convergence of the proposed heuristic algorithm for updating $\lambda$. The simulation conditions are the same as in Section IV.B.1 only with the difference that the maximum packet delay of VoIP service is set to 0.2s, 0.3s and 0.5s, respectively. (Denoted as Case 1 to 3) In Figure 3, all cases are depicted. Note that the values of $\lambda$ in different cases are normalized to show the convergence. The simulation shows that after 100 iterations the delay and throughput successfully converge in all cases.

## V. CONCLUSIONS

In this paper, we have studied subcarrier assignment in OFDMA networks and proposed a heuristic scheduling scheme to not only guarantee QoS provision for real-time services but also increase the total throughput for non-real-time applications. Based on the data-buffer state information, the channel state information and the QoS constraints, the scheduler provides a scheduling algorithm with given scheduling priorities that accounts both real-time and non-real-time users. And then we develop a heuristic for updating the priority for non-real-time users. Numerical results show that our scheme provides higher access data rate for non-real-time users in light load networks while guaranteeing necessary delay-control for real-time users and that the heuristic algorithm successfully converges under variable simulation conditions.